\date{January 2022}

\documentclass[11pt,reqno]{amsart}
\usepackage{latexsym,amsmath,amsfonts,amscd,amssymb}
\usepackage{graphics,color}
\usepackage{enumitem}
\usepackage{caption}
\usepackage{breqn}

\theoremstyle{plain}
\newtheorem{theorem}{Theorem}[section]

\theoremstyle{definition}
\newtheorem{definition}[theorem]{Definition}

\usepackage{hyperref}

\usepackage{tabularx}

\title{Vanna-Volga pricing for single and double barrier  FX options  }


\author[J. Mart\'{\i}n-Ovejero]{J. Mart\'{\i}n Ovejero}
\email{jmarovejero@gmail.com}

\begin{document}

\begin{abstract}
In this paper we provide a unified treatment of the Vanna-Volga pricing technique. We derive the value of single and double barriers FX options, as well as closed formulas for the Delta, Vega, Vanna and Volga of those contracts.
\end{abstract}

\maketitle

\section{Introduction}\label{sec:introduction}
The Foreign Exchange (FX) market is the most liquid OTC financial market in the world, being, at the same time, the largest market for options. This phenomenology is mainly due to the fact that the FX market is the only non-stop trading market. Nevertheless, currencies are traded in OTC markets, therefore, disclosures are not mandatory, which implies some opaqueness. There is a whole plethora of financial FX derivatives traded, ranging from the simple vanilla options and first-generation exotic products  (options with a single barrier, options with a double barrier, options with a window time barrier, and digital options between others) to the most complex financial products which have no closed-form formulas and some of them are hybrid.\\\

Using the well-known Black-Scholes model \cite{BS}, it is possible to calculate, analytically, the pricing value of vanilla and first-generation exotics FX products, but, the obtained results are far away to match the market quotations. This mismatching is a consequence of the unreasonable assumption that in the highly volatile FX market the foreign/domestic interest rates and the FX-spot volatility are constant through the lifetime of the financial product. More accurate models assume that volatilities evolution follows a stochastic process that depends on empirical observations. Nevertheless, for options with a short maturity (less than 1 year), interest rates can generally be assumed to be constant.\\\

From a theoretical point of view, stochastic volatility models allow explaining the volatility smile, but, as a drawback, great computational efforts must be made together with a calibration process to replicate the market dynamics. This problem has led to some clever pricing techniques that are fast and easy to implement. One of these methods is the so-called Vanna-Volga method, firstly described in \cite{Lipton}, where authors applied the method to the pricing of double no-touch options. After that, a systematic formulation of the method was described in \cite{Mercurio}, and some corrections handling the pricing inconsistencies can be found in \cite{bossens}. Roughly speaking, the Vanna-Volga method consists of adjusting the theoretical value of an option obtained from the Black Scholes model, adding the cost of the smile of a portfolio that hedges three main risks associated with the volatility of the option: the Vega, the Vanna, and the Volga. As it is well known, the Vega of a financial product is the sensitivity of the value of the contract with respect to the volatility of the underlying asset. By contrast, the Vanna  (resp. Volga) of a derivative is a second-order Greek that measures the sensitivity of the contract's Delta (resp. Vega) with respect to the volatility.\\\

The aim of this paper is to present a unified treatment of the Vanna-Volga pricing technique for first-gen exotic options, including single and double barrier contracts. We will derive closed formulas for the Greeks of those products that can be used by the working quant in the industry. It is important to remark that in the literature the Vanna-Volga technique is restricted to the FX market, nevertheless, we will explain how this technique can be easily adapted to any other market with high liquidity. 
\subsection*{Notation and conventions}
\begin{itemize}
	\item $S_t$ is the price of the underlying asset at time $t$.
	\item $K$ is the strike of an option.
	\item $n(x)=\frac{1}{\sqrt{2\pi}}e^{-\frac{x^2}{2}}$ is the standarized normal density function. 
	\item $\mathcal{N}(z)=\int_{-\infty}^{z}n(x) dx$ is the cumulative normal distribution function.
	\item $B,H,L$ are barriers.
	\item $r_d$ is the domestic risk free rate.
	\item $r_f$ is the foreign risk free rate.
	\item $T$ is the time to maturity (calculated according to the day count convention).
	\item $\mu$ is the drift.
	\item $\sigma$ is the volatility.
	\item $W$ is a Brownian motion.
	\item If $A$ is a subset of a set $X$, $\mathbb{I}_{A}:X\rightarrow\{0,1\}$ is the characteristic function of the subset $A$.
	\item $\text{erf}(z)$ is the error function.
	\item $\text{erfc}(z)=1-\text{erf}(z)$ is the complementary error function.
\end{itemize}

\section{Pricing Options with a single barrier}\label{sec:character}
\renewcommand{\arraystretch}{1.5}
Although pricing options with a single barrier is a well known  topic (see \cite{merton,rubinstein,haugh}), closed formulas for European options are disperse in the literature. Due to the fact that the Vanna-Volga pricing recipe involves the Black-Scholes price of the option, it is convenient for us to have summarized all the different formulas so, for the sake of completeness, they are included. Roughly speaking,  single barriers can be categorized in two groups: knock in and knock out barriers. The knock-out feature means that the option become worthless when the underlying asset price  crosses a predefined barrier. In contrast, options with knock in barriers are worthless until the underlying asset prices crosses a predefined barrier. Both barriers fall in two different classes: up and down. For instance, a down and out option has the barrier below the initial asset price and knocks out (become worthless) if the asset price falls bellow the barrier. Following the same reasoning  we have four different types of single barriers: up and out, up and in, down and out and down and in. In addition, each of those barriers can be reversed, meaning that the barrier is set in the money, rather than out the money. Therefore, we obtain sixteen different possible combinations, depending on the type of the option (call or put), the direction of the barrier (up and down), the type of the barrier (KI or KO) and if the barrier is standard or reversed.\\\

We restrict ourselves to options on one underlying asset without a pre-specified cash rebate paid out if the option has not been knocked in or has been knocked out during its lifetime. We assume that the price of the underlying asset  follows a Geometric Brownian Motion (GBM)
\begin{equation}
dS=\mu S dt+\sigma S dW
\end{equation}
where $\mu$ is the expected instantaneous rate of return on the underlying asset, $\sigma$ is the instantaneous volatility of the rate of return and $W$ is a Wiener process. Both, the risk free rate and the volatility, are assumed to be constant during the lifetime of the option. Let us observe, that in the particular case of FX-options
\begin{equation}
\mu=r_d-r_f
\end{equation}
where $r_d$ (resp. $r_d$) denotes the domestic (resp. foreign) interest rate.  In 1983, Garman and Holhagen (\cite{garman}) extended the BS model to cope with the existence of two interest rates. They obtained the following result.
\begin{theorem}
	Let $O$ be an FX-option with $S_0$ being the current spot rate, $K$ the strike price, and $T$ the time to maturity. Then, the price of the option is computed as
	\begin{equation}\label{bs-fx}\phi\Big(S_0 e^{-r_f T}\mathcal{N}(d_1)-K e^{-r_d T}\mathcal{N}(d_2)\Big)\end{equation}
	where
	\begin{align*}
	&d_1=\frac{\ln(\frac{S_0}{K})+(r_d-r_f+\sigma^2/2)T}{\sigma\sqrt T}\hspace{1cm}d_2=d_1-\sigma\sqrt{T}\\ \\
	&\phi=\begin{cases}
	\hspace{0.3cm}1 \hspace{1cm}\text{if $O$ is a call}\\
	-1\hspace{1cm}\text{if $O$ is a put}
	\end{cases}
	\end{align*}
\end{theorem}
The formulas developed in \cite{merton} for European options are valid for those options with only one interest rate. The extension of the aforementioned formulas in the case of FX-options can be found in \cite{haugh}. Those formulas use the following set of common parameters
\begin{equation}
\begin{aligned}\label{parameters}
&\mathcal{A}=\phi\Big(S_t e^{-r_f T}\mathcal{N}(\phi x_1)-K e^{-r_d T}\mathcal{N}(\phi(x_1-\sigma\sqrt{T})\Big)\\
&\mathcal{B}=\phi\Big(S_t e^{-r_f T}\mathcal{N}(\phi x_2)-K e^{-r_d T}\mathcal{N}(\phi(x_2-\sigma\sqrt{T})\Big)\\
&\mathcal{C}=\phi\Big[S_te^{-r_fT}(\frac{B}{S_t})^{2(\alpha+1)}\mathcal{N}(\eta y_1)-Ke^{-r_d T}(\frac{B}{S_t})^{2\alpha)}\mathcal{N}(\eta(y_1-\sigma\sqrt{T}))\Big]\\
&\mathcal{D}=\phi\Big[S_te^{-r_fT}(\frac{B}{S_t})^{2(\alpha+1)}\mathcal{N}(\eta y_2)-Ke^{-r_d T}(\frac{B}{S_t})^{2\alpha)}\mathcal{N}(\eta(y_2-\sigma\sqrt{T}))\Big]\\
&\alpha=\frac{r_d-r_f-\frac{\sigma^2}{2}}{\sigma^2}
\end{aligned}
\end{equation}
where
\begin{equation}
\nonumber
\begin{aligned}
&x_1=\frac{\ln(\frac{S_t}{K})}{\sigma\sqrt {T}}+(1+\alpha)\sigma\sqrt{T}
\hspace{1.2cm}x_2=\frac{\ln(\frac{S_t}{B})}{\sigma\sqrt {T}}+(1+\alpha)\sigma\sqrt{T}\\ \\ &y_1=\frac{\ln(\frac{B^2}{S_tK})}{\sigma\sqrt{T}}+(1+\alpha)\sigma\sqrt{T}\hspace{1cm} y_2=\frac{\ln(\frac{B}{S_t})}{\sigma\sqrt{T}}+(1+\alpha)\sigma\sqrt{T}\\ \\
&\phi=\begin{cases}
\hspace{0.3cm}1 \hspace{1cm}\text{if $O$ is a call}\\
-1\hspace{1cm}\text{if $O$ is a put}
\end{cases}\hspace{1cm }\eta=\begin{cases}
\hspace{0.3cm}1 \hspace{1cm}\text{if $B$ is a lower barrier}\\
-1\hspace{1cm}\text{if $B$ is an upper barrier}
\end{cases}
\end{aligned}
\end{equation}
\begin{theorem}Let $O$ be an FX-option with $S_0$ being the current spot rate, $K$ the strike price,  $T$ the time to maturity and $B$ the level barrier. Let
	\begin{equation}
	v(t,x)=e^{-r_d T}\mathbb{E}[F(S_T)\mid S_t=x]=e^{-r_d T}\mathbb{E}[F(S_t e^{(r_d-r_f-\sigma^2/2)T+\sigma\sqrt{T}Z})]
	\end{equation}
	the value of the derivative with payoff $F$ at time $t$ if the spot price is $S_t=x$, where $Z$ is the random variable representing the continuous returns.  Then, the values of $v(t,S_t)$ depending on the type of option, and on the type on the barrier, are summarized in the following table
	
	\begin{center}
		\captionof{table}{Pricing one single barriers}\label{table-price-bar}
		
		\begin{tabular}{ | c | c | c | c| c |}
			
			\hline
			Option & $\phi$ & $\eta$ & Inequality & Value \\ \hline
			Up and In Call&+1&-1& $K>B$ & $\mathcal A$ \\ \hline
			Down and In Call&+1&+1& $K > B$ & $\mathcal C$ \\ \hline
			Up and  Out Call&+1&-1& $K> B$ & $0$ \\ \hline
			Down and Out Call&+1&+1& $K>B$ & $\mathcal A-\mathcal C$ \\ \hline
			Reverse Up and In Call&+1&-1& $K\leq B$ & $\mathcal B-\mathcal C+\mathcal D$ \\ \hline
			Reverse Down and In Call&+1&+1& $K\leq B$ & $\mathcal A-\mathcal B+\mathcal D$ \\ \hline
			Reverse Up and Out Call&+1&-1& $K \leq B$ &$\mathcal A-\mathcal B+\mathcal C-\mathcal D$ \\ \hline
			Reverse Down and Out Call&+1&+1& $K\leq B$ & $\mathcal B-\mathcal D$ \\ \hline
			Up and In Put&-1&-1& $K\leq B$ & $\mathcal C$ \\ \hline
			Down and In Put&-1&+1& $K\leq B$ & $\mathcal A$ \\ \hline
			Up and  Out Put&-1&-1& $K\leq B$ & $\mathcal A-\mathcal C$ \\ \hline
			Down and Out Put&-1&+1& $K\leq B$ & $0$ \\ \hline
			Reverse Up and In Put&-1&-1& $K> B$ & $\mathcal A-\mathcal B+\mathcal D$ \\ \hline
			Reverse Down and In Put&-1&+1& $K> B$ & $\mathcal B-\mathcal C+\mathcal D$ \\ \hline
			Reverse Up and Out Put&-1&-1& $K>B$ & $\mathcal B-\mathcal D$ \\ \hline
			Reverse Down and Out Put&-1&+1& $K>B$ & $\mathcal A-\mathcal B+\mathcal C-\mathcal D$ \\ \hline

		\end{tabular}

	\end{center}
	where $\mathcal A,\mathcal B,\mathcal C,\mathcal D$ are the set of parameters (depending on $S_t$) defined in \eqref{parameters}
\end{theorem}
\begin{proof}
	The idea of the proof goes as follows. We have assumed before that the option's underlying follows a GBM. It is standard to compute closed form solutions for different options types with payoff $F(S_t)$ at maturity via
	
	$$v(t,x)=e^{-r_d T}\mathbb{E}[F(S_T)\mid S_t=x]=e^{-r_d T}\mathbb{E}[F(S_t e^{(r_d-r_f-\sigma^2/2)T+\sigma\sqrt{T}Z})]$$
	
	where $Z$ is the random variable representing the continuous returns which is modelled as standard normal in the BS model. Therefore
	
	\begin{equation}\label{vxt}
	v(t,x):=e^{-r_d T}\int_{-\infty}^{\infty}F\Big(S_te^{(r_d-r_f-\sigma^2/2)T+\sigma\sqrt{T}z}\Big)n(z)dz\end{equation}
	
	To price knock-out options, we consider the payoff
	
	\begin{equation}\label{payoff}[\phi(S_T-K)]^{+}\mathbb{I}_{\{\min_{t\in[0,T]}(\eta S_t)>\eta B\}}
	\end{equation}
	
	where
	
	$$[\phi(S_T-K)]^{+}=\max(\phi(S_t-K),0)$$
	
	For the knock-in options we can use the In-Out parity for barrier options in the absence of rebates.  This parity means that being long a knock-out option, and being long a knock-in option with the same features (option type, barrier level, strike) is equivalent to owning a vanilla option independently from the behaviour of the spot with respect to the barrier level. Obtaining the value of a barrier option in the BS-model is reduced to knowing the joint density $f(x,y)$ for a Brownian motion with drift and its running extremum ($\eta=+1$ and $\eta=-1$ for a minimum) of the pair
	
	$$(W(T)+(\frac{r_d-r_f}{\sigma}-\frac{\sigma^2}{2})T, \eta\min_{0\leq t\leq T}[\eta(W(t)+(\frac{r_d-r_f}{\sigma}-\frac{\sigma^2}{2})t)]$$
	
	This joint distribution can be found in \cite[Ch. 7]{shreve} and can be expressed as
	
	\begin{equation}\label{joint}
	\begin{aligned}
	f(x,y)=-\eta e^{(\frac{r_d-r_f}{\sigma}-\frac{\sigma^2}{2})x-\frac{1}{2}(\frac{r_d-r_f}{\sigma}-\frac{\sigma^2}{2})^2 T}\frac{2(2y-x)}{T\sqrt{2\pi T}}e^{-\frac{(2y-x)^2}{2T}}\\
	\eta y\leq \min (0,\eta x)
	\end{aligned}
	\end{equation}
	
	Using \eqref{vxt}, \eqref{payoff} and \eqref{joint}, the value of a barrier option can be written as the following integral
	
	\begin{equation}\label{inte}
	e^{-r_d T}\int_{x=-\infty}^{x=\infty}\int_{\eta y\leq \min(0,\eta,x)}[\phi(S_0 e^{\sigma_x}-K)]^{+}\mathbb{I}_{\{\eta y<\eta\frac{1}{\sigma}\ln\frac{B}{S_0}\}}f(x,y)dydx
	\end{equation}
	In \cite[Ch. 7]{shreve} all the details needed to evaluate \eqref{inte} are provided. That integral produces the four different terms listed in \eqref{parameters}.
\end{proof}
\section{Pricing Options with Double Barriers}
This section is devoted to the pricing formulas for FX-options equipped with double barriers. As the name suggests, a double barrier option is an exotic option whose payoff is determined given two barrier levels: an upper one and a lower one. The valuation of these financial derivatives is really problematic. Different techniques can be found throughout the literature, from numerical methods such as the finite difference method, to econometric models like the binomial and trinomial model. In this article we will follow the techniques developed by Ikeda and Kunitomo in \cite{kunitomo}. In the aforementioned paper, the authors develop a common framework to evaluate option contracts with two curved boundaries. The key point of the paper is the generalization of the Levy formula on the
Brownian motion by T. W. Anderson in sequential analysis. The  general pricing formulae for options with
boundaries are expressed as infinite series, but after carrying out a careful numerical study, the authors find that the convergence of the series is very fast, and most of the terms tend to zero very quickly. The following theorems are a direct application of the results obtained in \cite{kunitomo} by setting the curvature of both barriers equal to zero.
\begin{theorem}[\cite{kunitomo}] The valuation of a double knock out call option, with strike $K$, initial spot price $S$ and barriers $L<K<U$ is
	\begin{equation}\label{pricedcall}
	\begin{aligned}
	KOKOC&=e^{-r_f T}S\sum_{n=-\infty}^{\infty}\Big[\Big(\frac{U^n}{L^n}\Big)^{\alpha}[\mathcal{N}(d_1)-\mathcal{N}(d_2)]-\Big(\frac{L^{n+1}}{U^{n}S}\Big)^{\alpha}[\mathcal{N}(d_3)-\mathcal{N}(d_4)]\Big]\\
	&-e^{-r_d T}K\sum_{n=-\infty}^{\infty}\Big[\Big(\frac{U^{n}}{L^{n}}\Big)^{\alpha-2}[\mathcal{N}(d_1-\sigma\sqrt{T})-\mathcal{N}(d_2-\sigma\sqrt{T})]\\
	&-\Big(\frac{L^{n+1}}{U^{n}S}\Big)^{\alpha-2}[\mathcal{N}(d_3-\sigma\sqrt{T})-\mathcal{N}(d_4-\sigma\sqrt{T})]\Big]
	\end{aligned}
	\end{equation}
	
	where
	
	\begin{equation}
	\nonumber
	\begin{aligned}
	&d_1=\frac{\ln(\frac{SU^{2n}}{KL^{2n}})+(b+\frac{\sigma^2}{2})T}{\sigma\sqrt{T}}
	\hspace{1.2cm}d_2=\frac{\ln(\frac{SU^{2n}}{UL^{2n}})+(b+\frac{\sigma^2}{2})T}{\sigma\sqrt{T}}\\ \\ &d_3=\frac{\ln(\frac{L^{2n+2}}{KSU^{2n}})+(b+\frac{\sigma^2}{2})T}{\sigma\sqrt{T}}\hspace{1cm} d_4=\frac{\ln(\frac{L^{2n+2}}{SU^{2n+1}})+(b+\frac{\sigma^2}{2})T}{\sigma\sqrt{T}}\\ \\
	&\alpha=\frac{2b}{\sigma^2}+1 \hspace{4cm}b=r_d-r_f
	\end{aligned}
	\end{equation}
	
	The valuation of a double knock out put option, with strike $K$, initial spot price $S$ and barriers $L<K<U$ is
	
	\begin{equation}\label{pricedput}
	\begin{aligned}
	&KOKOP=e^{-r_d T}K\sum_{n=-\infty}^{\infty}\Big[\Big(\frac{U^{n}}{L^{n}}\Big)^{\alpha-2}[\mathcal{N}(y_1-\sigma\sqrt{T})-\mathcal{N}(y_2-\sigma\sqrt{T})]\\
	&-\Big(\frac{L^{n+1}}{U^{n}S}\Big)^{\alpha-2}[\mathcal{N}(y_3-\sigma\sqrt{T})-\mathcal{N}(y_4-\sigma\sqrt{T})]\Big]\\
	&-e^{-r_f T}S\sum_{n=-\infty}^{\infty}\Big[\Big(\frac{U^n}{L^n}\Big)^{\alpha}[\mathcal{N}(y_1)-\mathcal{N}(y_2)]-\Big(\frac{L^{n+1}}{U^{n}S}\Big)^{\alpha}[\mathcal{N}(y_3)-\mathcal{N}(y_4)]\Big]
	\end{aligned}
	\end{equation}
	where
	
	\begin{equation}
	\nonumber
	\begin{aligned}
	&y_1=\frac{\ln(\frac{SU^{2n}}{L^{2n+1}})+(b+\frac{\sigma^2}{2})T}{\sigma\sqrt{T}}
	\hspace{1.2cm}y_2=\frac{\ln(\frac{SU^{2n}}{KL^{2n}})+(b+\frac{\sigma^2}{2})T}{\sigma\sqrt{T}}\\ \\ &y_3=\frac{\ln(\frac{L^{2n+2}}{LSU^{2n}})+(b+\frac{\sigma^2}{2})T}{\sigma\sqrt{T}}\hspace{1cm} y_4=\frac{\ln(\frac{L^{2n+2}}{KSU^{2n}})+(b+\frac{\sigma^2}{2})T}{\sigma\sqrt{T}}\\ \\
	&\alpha=\frac{2b}{\sigma^2}+1 \hspace{4cm}b=r_d-r_f
	\end{aligned}
	\end{equation}
\end{theorem}
The numerical study carried out in \cite{kunitomo} suggests that it suffices to calculate  the leading five terms for most cases, therefore, when implementing the code, we will compute the previous analytical series from $n=-5$ to $n=5$. Once that we have computed the valuation of double knock-out calls and puts, we are going to price other double barrier products through  replicating  portfolios. For the sake of consistency, let us price first the double knock-in FX-options.
\begin{theorem}\hfill\label{t:kiki}
	\begin{enumerate}
		\item The price of a double knock-in call FX-option  with barriers $L<K<U$ is
		\begin{equation}
		\Big(S_0 e^{-r_f T}\mathcal{N}(\overline{d}_1)-K e^{-r_d T}\mathcal{N}(\overline{d}_2)\Big)-KOKOC(L,U)
		\end{equation}
		
		\item The price of a double knock-in put FX-option  with barriers $L<K<U$ is
		\begin{equation}
		\Big(K e^{-r_d T}\mathcal{N}(\overline{d}_2)+S_0 e^{-r_f T}\mathcal{N}(\overline{d}_1)\Big)-KOKOP(L,U)
		\end{equation}
		being $KOKOC(L,U)$ (resp. $KOKOP(L,U)$)  the price of a double knock-out FX call (resp. put) with barriers $L$ and $U$ (see \eqref{pricedcall} and \eqref{pricedput}), and where
		\begin{align*}
		&\overline{d}_1=\frac{\ln(\frac{S_0}{K})+(r_d-r_f+\sigma^2/2)T}{\sigma\sqrt T}\hspace{1cm}\overline{d}_2=\overline{d}_1-\sigma\sqrt{T}
		\end{align*}
	\end{enumerate}
\end{theorem}
\begin{proof}
	$(1)$  The term $\Big(K e^{-r_d T}\mathcal{N}(\overline{d}_2)+S_0 e^{-r_f T}\mathcal{N}(\overline{d}_1)\Big)$ is the predicted value in the Black-Scholes model (see \eqref{bs-fx}) for a vanilla call FX-option. Let us proof that the payoff of a double knock-in option is the same as being long a vanilla call with the same parameters, and being short a double knock-out FX-option with the same barriers. We have to consider three different scenarios.
	\begin{enumerate}
		\item The spot prices does not touch either of the barriers, so, the final payoff for the double knock-in call is $0$.  The payoff of the vanilla call and the payoff of the double knock-out call is the same.
		\item If the spot hist the upper barrier, then, the payoff for the double knock-in option is the same as a vanilla call. In that case, since the spot hits the upper barrier, the double knock-out call is terminated.
		\item  If the spot hits the lower barrier, the payoff $0$. Again, the double knock out call is terminated.
	\end{enumerate}
	Through this replicating portfolio, we conclude that $KIKI=Vanilla-KOKO$. An analogous reasoning allow us to prove $(2)$, which is the parallel case for put options.
\end{proof}
One of the most popular first generation exotic FX-options are the KIKO options.  A KIKO option is an option equipped with two barriers, one knock-in and one knock-out. In this product, the knock in barrier must be hit in order to active the underling vanilla option. In addition, the knock out barrier is valid throughout the lifetime of the option, which causes the option to be terminated if it is hit. Let us recall that even the underlying option is activated, it can still be extinguished at any time until expiration. The two barriers can be a combination of regular and reverse barriers. Nevertheless,  if both barriers are on the same side of the underlying asset, the knock in barrier must be defined between the underlying and the knock out barrier, otherwise, the payoff will always be zero.  As we did in the proof of Theorem \eqref{t:kiki}, we can use the portfolio replicating technique to price all the possible combinations of KIKO options as it showed in the following table. For the sake of clarity, let us denote by $B_I$ the knock-in barrier and by $B_O$ the knock-out barrier, and let us suppose that all the options have the same parameters: maturity time, strike, volatility, domestic and foreign interest rates.

\begin{center}
	\captionof{table}{Pricing KIKO  barriers}\label{table-price-2bar}
	
	\begin{tabular}{ | c | c | c |}
		
		\hline
		Option & Inequality & Replicating portfolio \\ \hline
		Call&$B_I<K\leq B_O$&  $RUOC(B_O)-KOKOC(B_I,B_O)$ \\ \hline
		Call& $K\leq B_I< B_O$& $RUIC(B_I)-RUPIC(B_O)$\\ \hline
		Call& $K\leq B_O< B_I$& $0$\\ \hline
		Call& $ B_I< B_O<K$& $0$\\ \hline
		Call& $B_O< B_I<K$& $DIC(B_I)-DIC(B_O)$\\ \hline
		Call& $B_O< K\leq B_I$& $RUIC(B_I)-DIC(B_O)$\\ \hline
		Put&$B_I\leq K< B_O$&  $RDIP(B_I)-UPIP(B_O)$ \\ \hline
		Put& $K< B_I< B_O$& $UPIP(B_I)-UPIP(B_O)$\\ \hline
		Put& $K\leq B_O< B_I$& $0$\\ \hline
		Put& $ B_I< B_O\leq K$& $0$\\ \hline
		Put& $B_O< B_I\leq K$& $RDOP(B_O)-KOKOP(B_I,B_0)$\\ \hline
		Put& $B_O\leq K< B_I$& $UIP(B_I)-RDIP(B_O)$\\ \hline
		
	\end{tabular}
\end{center}
\section{Computation of Greeks}
The aim of this section is to provide a complete set of explicit formulas for the Delta,  Vega,  Vanna and  Volga for all of the different first-gen exotics  explained in the  previous sections. Although those estimates are fairly easy to derive from the pricing formulas we have developed before, we think it is convenient for the working quant to have them summarized. Since we have obtained the price of all single barrier options from the parameters $\mathcal{A},\mathcal{B}, \mathcal C$ and $\mathcal D$, along with the fact that the derivative is a linear operator,  it is enough to calculate the aforementioned Greeks for $\mathcal{A},\mathcal{B}, \mathcal C$ and $\mathcal D$ to obtain the Greeks of a single barrier option. The case of double barriers is reduced to the case of single barriers by using the replicating portfolio technique. The only extra computation needed is to derive the Greeks for $KOKO$ options (\eqref{pricedcall} and \eqref{pricedput}) truncating the series to $n=5$. The Greeks for the $KIKI$ option can be obtained through the relationship $KIKI=Vanilla -KOKO$. This last step is left as an easy exercise to the reader. 

\begin{definition}
	The Greeks are the quantities representing the sensitivity of the price of derivatives with respect to a change in underlying parameters on which the value of the derivative is dependent.\\\
	
	Let $V$ denotes  the value of the derivative, then, we define the Delta, as the sensitivity of $V$ with respect to a change on the underlying instrument's price $S$
	\begin{equation}
	\Delta=\frac{\partial V}{\partial S}
	\end{equation}
	We define the Vega as the sensitivity of $V$ with respect to a change of the volatility
	\begin{equation}
	Vega=\frac{\partial V}{\partial \sigma}
	\end{equation}
	The Vanna is the sensitivity of the Delta with respect to a change in the volatility
	\begin{equation}
	Vanna=\frac{\partial \Delta}{\partial \sigma}
	\end{equation}
	The Volga measures the second order sensitivity to volatility, that is
	\begin{equation}
	Volga=\frac{\partial^2 V}{\partial \sigma^2}
	\end{equation}
\end{definition}
\subsection{Computations of Greeks for single barriers}
In this section we summarize all the different Greeks for the parameters $\mathcal{A},\mathcal{B}, \mathcal C$ and $\mathcal D$.
%
%

\begin{equation}
\nonumber
\begin{aligned}
	&\Delta(\mathcal{A})=
	\frac{1}{2} \phi e^{-r_f T} (\text{erf}(\frac{\phi (2 \log (\frac{S}{K}))+T (2 r_d-2 r_f+\sigma ^2)}{2 \sqrt{2} \sigma  \sqrt{T}})+1)
\\\\
&Vega(\mathcal{A})=
\frac{K \sqrt{T}  }{\sqrt{2 \pi }}(\frac{S}{K})^{-\frac{r_d}{\sigma ^2}+\frac{r_f}{\sigma ^2}+\frac{1}{2}} \exp (-\frac{4 \log ^2(\frac{S}{K})+T^2 (4 r_d^2+r_d (4 \sigma ^2-8 r_f)+(2 r_f+\sigma ^2)^2)}{8 \sigma ^2 T})
\\\\
&Vanna(\mathcal A)=\\ &\frac{ (\frac{S}{K})^{-\frac{r_d}{\sigma ^2}+\frac{r_f}{\sigma ^2}-\frac{1}{2}}}{2 \sqrt{2 \pi } \sigma ^2 \sqrt{T}} (T (-2 r_d+2 r_f+\sigma ^2)-2 \log (\frac{S}{K})) e^{-\frac{4 \log ^2(\frac{S}{K}))+T^2 (4 r_d^2+r_d (4 \sigma ^2-8 r_f))+(2 r_f+\sigma ^2)^2)}{8 \sigma ^2 T})}
\end{aligned}
\end{equation}

\begin{equation}
\nonumber
\begin{aligned}
&Volga(\mathcal A)=\\ &\frac{1}{8 \sqrt{2 \pi } \sigma ^5 T^{3/2}} (-16 K \sigma ^2 T (\log (\frac{S}{K})+T (r_d-r_f)) \exp (-\frac{(T (-2 r_d+2 r_f+\sigma ^2)-2 \log (\frac{S}{K}))^2}{8 \sigma ^2 T}-r_d T)\\&+16 \sigma ^2 S T (\log (\frac{S}{K})+T (r_d-r_f)) \exp (-\frac{(2 \log (\frac{S}{K})+T (2 r_d-2 r_f+\sigma ^2))^2}{8 \sigma ^2 T}-r_f T)\\ &+K (2 \log (\frac{S}{K})+T (2 r_d-2 r_f-\sigma ^2)) (2 \log (\frac{S}{K})+T (2 r_d-2 r_f+\sigma ^2))^2\times \\ &\times\exp (-\frac{(T (-2 r_d+2 r_f+\sigma ^2)-2 \log (\frac{S}{K}))^2}{8 \sigma ^2 T}-r_d T)-S (2 \log (\frac{S}{K})+T (2 r_d-2 r_f-\sigma ^2))^2\times\\&\times (2 \log (\frac{S}{K})+T (2 r_d-2 r_f+\sigma ^2)) \exp (-\frac{(2 \log (\frac{S}{K})+T (2 r_d-2 r_f+\sigma ^2))^2}{8 \sigma ^2 T}-r_f T))
\\\\
&\Delta(\mathcal{B})=\\&
\phi  (\frac{1}{2} e^{-r_f T} (\text{erf}(\frac{\phi  (2 \log (\frac{S}{B})+T (2 r_d-2 r_f+\sigma ^2))}{2 \sqrt{2} \sigma  \sqrt{T}})+1)-\\ &\frac{K \phi  \exp (-\frac{(T (-2 r_d+2 r_f+\sigma ^2)-2 \log (\frac{S}{B}))^2}{8 \sigma ^2 T}-r_d T)}{\sqrt{2 \pi } \sigma  S \sqrt{T}}+\frac{\phi  \exp (-\frac{(2 \log (\frac{S}{B})+T (2 r_d-2 r_f+\sigma ^2))^2}{8 \sigma ^2 T}-r_f T)}{\sqrt{2 \pi } \sigma  \sqrt{T}})
\\\\
&Vega(\mathcal{B})=\\&
\frac{1}{2 \sqrt{2 \pi } \sigma ^2 \sqrt{T}}( K (2 \log (\frac{S}{B})+T (2 r_d-2 r_f+\sigma ^2)) \exp (-\frac{(T (-2 r_d+2 r_f+\sigma ^2)-2 \log (\frac{S}{B}))^2}{8 \sigma ^2 T}-r_d T)\\&+ S (T (-2 r_d+2 r_f+\sigma ^2)-2 \log (\frac{S}{B})) \exp (-\frac{(2 \log (\frac{S}{B})+T (2 r_d-2 r_f+\sigma ^2))^2}{8 \sigma ^2 T}-r_f T))
\end{aligned}
\end{equation}

\begin{equation}
\nonumber
\begin{aligned}
&Vanna(\mathcal{B})=\\&
\frac{1}{4 \sqrt{2 \pi } \sigma ^4 S T^{3/2}} (-K (8 T (r_d-r_f) \log (\frac{S}{B})+4 \log ^2(\frac{S}{B})+T^2 (4 r_d^2-8 r_d r_f+4 r_f^2-\sigma ^4))\times\\&\times \exp (-\frac{(T (-2 r_d+2 r_f+\sigma ^2)-2 \log (\frac{S}{B}))^2}{8 \sigma ^2 T}-r_d T)\\&+4 K \sigma ^2 T \exp (-\frac{(T (-2 r_d+2 r_f+\sigma ^2)-2 \log (\frac{S}{B}))^2}{8 \sigma ^2 T}-r_d T)+\\&S (8 T (r_d-r_f) \log (\frac{S}{B})+4 \log ^2(\frac{S}{B})+T^2 (4 r_d^2-8 r_d r_f+4 r_f^2-\sigma ^4))\times\\&\times \exp (-\frac{(2 \log (\frac{S}{B})+T (2 r_d-2 r_f+\sigma ^2))^2}{8 \sigma ^2 T}-r_f T)\\&-4 \sigma ^2 S T \exp (-\frac{(2 \log (\frac{S}{B})+T (2 r_d-2 r_f+\sigma ^2))^2}{8 \sigma ^2 T}-r_f T)\\&+2 \sigma ^2 S T (T (-2 r_d+2 r_f+\sigma ^2)-2 \log (\frac{S}{B})) \exp (-\frac{(2 \log (\frac{S}{B})+T (2 r_d-2 r_f+\sigma ^2))^2}{8 \sigma ^2 T}-r_f T))
\end{aligned}
\end{equation}
\begin{equation}
\nonumber
\begin{aligned}
&Volga(\mathcal{B})=\\&
\frac{1}{8 \sqrt{2 \pi } \sigma ^5 T^{3/2}}(-16 K \sigma ^2 T (\log (\frac{S}{B})+T (r_d-r_f)) \exp (-\frac{(T (-2 r_d+2 r_f+\sigma ^2)-2 \log (\frac{S}{B}))^2}{8 \sigma ^2 T}-r_d T)\\&+K (2 \log (\frac{S}{B})+T (2 r_d-2 r_f-\sigma ^2)) (2 \log (\frac{S}{B})+T (2 r_d-2 r_f+\sigma ^2))^2 \times\\&\times \exp (-\frac{(T (-2 r_d+2 r_f+\sigma ^2)-2 \log (\frac{S}{B}))^2}{8 \sigma ^2 T}-r_d T)\\& +16 \sigma ^2 S T (\log (\frac{S}{B})+T (r_d-r_f)) \exp (-\frac{(2 \log (\frac{S}{B})+T (2 r_d-2 r_f+\sigma ^2))^2}{8 \sigma ^2 T}-r_f T)\\&-S (2 \log (\frac{S}{B})+T (2 r_d-2 r_f-\sigma ^2))^2 (2 \log (\frac{S}{B})+T (2 r_d-2 r_f+\sigma ^2)) \times\\&\times \exp (-\frac{(2 \log (\frac{S}{B})+T (2 r_d-2 r_f+\sigma ^2))^2}{8 \sigma ^2 T}-r_f T))
\\\\
&\Delta(\mathcal{C})=\\&
\frac{\phi}{2 \sigma ^2 S^2 \sqrt{T}}  (\frac{B}{S})^{\frac{2 (r_d-r_f)}{\sigma ^2}-1} (2 B^2 \sqrt{T} (r_f-r_d) e^{-r_f T} \text{erf}(\frac{\eta  (2 \log (\frac{B^2}{K S})+T (2 r_d-2 r_f+\sigma ^2))}{2 \sqrt{2} \sigma  \sqrt{T}})\\&+K S \sqrt{T} e^{-r_d T} (2 r_d-2 r_f-\sigma ^2) \text{erfc}(\frac{\eta  (T (-2 r_d+2 r_f+\sigma ^2)-2 \log (\frac{B^2}{K S}))}{2 \sqrt{2} \sigma  \sqrt{T}})\\&+\sqrt{\frac{2}{\pi }} B^2 \eta  \sigma  (-\exp (-\frac{\eta ^2 (2 \log (\frac{B^2}{K S})+T (2 r_d-2 r_f+\sigma ^2))^2}{8 \sigma ^2 T}-r_f T))\\&+\sqrt{\frac{2}{\pi }} \eta  K \sigma  S \exp (-\frac{\eta ^2 (T (-2 r_d+2 r_f+\sigma ^2)-2 \log (\frac{B^2}{K S}))^2}{8 \sigma ^2 T}-r_d T)\\&-2 B^2 r_d \sqrt{T} e^{-r_f T}+2 B^2 r_f \sqrt{T} e^{-r_f T})
\\\\
&Vega(\mathcal C)=\\&
\frac{\phi}{4 B \sigma ^3}  (\frac{B}{S})^{\frac{2 (r_d-r_f)}{\sigma ^2}} (8 B^2 (r_f-r_d) e^{-r_f T} \log (\frac{B}{S}) (\text{erf}(\frac{\eta  (2 \log (\frac{B^2}{K S})+T (2 r_d-2 r_f+\sigma ^2))}{2 \sqrt{2} \sigma  \sqrt{T}})+1)\\&+8 K S (r_d-r_f) e^{-r_d T} \log (\frac{B}{S}) \text{erfc}(\frac{\eta  (T (-2 r_d+2 r_f+\sigma ^2)-2 \log (\frac{B^2}{K S}))}{2 \sqrt{2} \sigma  \sqrt{T}})\\&+\frac{\sqrt{\frac{2}{\pi }} B^2 \eta  \sigma  (T (-2 r_d+2 r_f+\sigma ^2)-2 \log (\frac{B^2}{K S})) \exp (-\frac{\eta ^2 (2 \log (\frac{B^2}{K S})+T (2 r_d-2 r_f+\sigma ^2))^2}{8 \sigma ^2 T}-r_f T)}{\sqrt{T}}\\&+\frac{\sqrt{\frac{2}{\pi }} \eta  K \sigma  S (2 \log (\frac{B^2}{K S})+T (2 r_d-2 r_f+\sigma ^2)) \exp (-\frac{\eta ^2 (T (-2 r_d+2 r_f+\sigma ^2)-2 \log (\frac{B^2}{K S}))^2}{8 \sigma ^2 T}-r_d T)}{\sqrt{T}})
\end{aligned}
\end{equation}
\begin{equation}
\nonumber
\begin{aligned}
&Vanna(\mathcal C)=\\&
\frac{(\frac{B}{S})^{\frac{2 (r_d-r_f)}{\sigma ^2}-1} \phi}{8 \sqrt{\pi } S^2 T^{3/2} \sigma ^5} (-8 e^{-r_d T} K \sqrt{\pi } S \sigma ^4 \text{erfc}(\frac{\eta  (T (\sigma ^2-2 r_d+2 r_f)-2 \log (\frac{B^2}{K S}))}{2 \sqrt{2} \sqrt{T} \sigma }) T^{3/2}\\&-16 e^{-r_d T} K \sqrt{\pi } (r_d-r_f) S (-\sigma ^2+2 r_d-2 r_f) \text{erfc}(\frac{\eta  (T (\sigma ^2-2 r_d+2 r_f)-2 \log (\frac{B^2}{K S}))}{2 \sqrt{2} \sqrt{T} \sigma }) \log (\frac{B}{S}) T^{3/2}\\&+16 B^2 e^{-r_f T} (r_d-r_f) \sigma ^2 (\text{erf}(\frac{\eta  (T (\sigma ^2+2 r_d-2 r_f)+2 \log (\frac{B^2}{K S}))}{2 \sqrt{2} \sqrt{T} \sigma })+1) \sqrt{\pi } T^{3/2}\\&+8 e^{-r_d T} K S \sigma ^2 (\sigma ^2-2 r_d+2 r_f) \text{erfc}(\frac{\eta  (T (\sigma ^2-2 r_d+2 r_f)-2 \log (\frac{B^2}{K S}))}{2 \sqrt{2} \sqrt{T} \sigma }) \sqrt{\pi } T^{3/2}\\&+16 B^2 e^{-r_f T} (r_f-r_d) \sigma ^2 (\text{erf}(\frac{\eta  (T (\sigma ^2+2 r_d-2 r_f)+2 \log (\frac{B^2}{K S}))}{2 \sqrt{2} \sqrt{T} \sigma })+1) \log (\frac{B}{S}) \sqrt{\pi } T^{3/2}\\&+16 B^2 e^{-r_f T} (r_d-r_f) (\sigma ^2+2 r_d-2 r_f) (\text{erf}(\frac{\eta  (T (\sigma ^2+2 r_d-2 r_f)+2 \log (\frac{B^2}{K S}))}{2 \sqrt{2} \sqrt{T} \sigma })+1) \log (\frac{B}{S}) \times\\& \times\sqrt{\pi } T^{3/2}-4 \sqrt{2} e^{-\frac{\eta ^2 (T (\sigma ^2-2 r_d+2 r_f)-2 \log (\frac{B^2}{K S}))^2}{8 T \sigma ^2}-r_d T} K S \eta  \sigma ^3 T\\&-16 \sqrt{2} e^{-\frac{\eta ^2 (T (\sigma ^2-2 r_d+2 r_f)-2 \log (\frac{B^2}{K S}))^2}{8 T \sigma ^2}-r_d T} K (r_d-r_f) S \eta  \sigma  \log (\frac{B}{S}) T\\&-2 \sqrt{2} B^2 e^{-\frac{\eta ^2 (T (\sigma ^2+2 r_d-2 r_f)+2 \log (\frac{B^2}{K S}))^2}{8 T \sigma ^2}-r_f T} \eta  \sigma  (\sigma ^2+2 r_d-2 r_f) (T (\sigma ^2-2 r_d+2 r_f)-2 \log (\frac{B^2}{K S})) T\\&+4 B^2 e^{-\frac{\eta ^2 (T (\sigma ^2+2 r_d-2 r_f)+2 \log (\frac{B^2}{K S}))^2}{8 T \sigma ^2}-r_f T} \eta  \sigma ^3 \sqrt{2} T+16 B^2 e^{-\frac{\eta ^2 (T (\sigma ^2+2 r_d-2 r_f)+2 \log (\frac{B^2}{K S}))^2}{8 T \sigma ^2}-r_f T}\times\\ &\times (r_d-r_f) \eta  \sigma  \log (\frac{B}{S}) \sqrt{2} T+2 e^{-\frac{\eta ^2 (T (\sigma ^2-2 r_d+2 r_f)-2 \log (\frac{B^2}{K S}))^2}{8 T \sigma ^2}-r_d T} K S \eta  \sigma  (\sigma ^2-2 r_d+2 r_f)\times\\&\times (T (\sigma ^2+2 r_d-2 r_f)+2 \log (\frac{B^2}{K S})) \sqrt{2} T+2 B^2 e^{-\frac{\eta ^2 (T (\sigma ^2+2 r_d-2 r_f)+2 \log (\frac{B^2}{K S}))^2}{8 T \sigma ^2}-r_f T} \eta  \sigma ^3\times \\&\times  (T (\sigma ^2-2 r_d+2 r_f)-2 \log (\frac{B^2}{K S})) \sqrt{2} T-\sqrt{2} e^{-\frac{\eta ^2 (T (\sigma ^2-2 r_d+2 r_f)-2 \log (\frac{B^2}{K S}))^2}{8 T \sigma ^2}-r_d T} K S \eta ^3 \sigma\times \\&\times  ((\sigma ^4-4 r_d^2-4 r_f^2+8 r_d r_f) T^2+8 (r_f-r_d) \log (\frac{B^2}{K S}) T-4 \log ^2(\frac{B^2}{K S}))\\& +B^2 e^{-\frac{\eta ^2 (T (\sigma ^2+2 r_d-2 r_f)+2 \log (\frac{B^2}{K S}))^2}{8 T \sigma ^2}-r_f T} \eta ^3 \sigma  ((\sigma ^4-4 r_d^2-4 r_f^2+8 r_d r_f) T^2+8 (r_f-r_d) \log (\frac{B^2}{K S}) T\\&-4 \log ^2(\frac{B^2}{K S})) \sqrt{2})
\end{aligned}
\end{equation}
\begin{equation}
\nonumber
\begin{aligned}
&Volga(\mathcal C)=\\&
\frac{(\frac{B}{S})^{\frac{2 (r_d-r_f)}{\sigma ^2}} \phi}{16 B \sqrt{\pi } T^{3/2} \sigma ^6} (-128 e^{-r_d T} K \sqrt{\pi } (r_d-r_f)^2 S \text{erfc}(\frac{\eta  (T (\sigma ^2-2 r_d+2 r_f)-2 \log (\frac{B^2}{K S}))}{2 \sqrt{2} \sqrt{T} \sigma }) \log ^2(\frac{B}{S}) T^{3/2}\\&-96 e^{-r_d T} K \sqrt{\pi } (r_d-r_f) S \sigma ^2 \text{erfc}(\frac{\eta  (T (\sigma ^2-2 r_d+2 r_f)-2 \log (\frac{B^2}{K S}))}{2 \sqrt{2} \sqrt{T} \sigma }) \log (\frac{B}{S}) T^{3/2}\\&+128 B^2 e^{-r_f T} (r_d-r_f)^2 (\text{erf}(\frac{\eta  (T (\sigma ^2+2 r_d-2 r_f)+2 \log (\frac{B^2}{K S}))}{2 \sqrt{2} \sqrt{T} \sigma })+1) \log ^2(\frac{B}{S}) \sqrt{\pi } T^{3/2}\\&+96 B^2 e^{-r_f T} (r_d-r_f) \sigma ^2 (\text{erf}(\frac{\eta  (T (\sigma ^2+2 r_d-2 r_f)+2 \log (\frac{B^2}{K S}))}{2 \sqrt{2} \sqrt{T} \sigma })+1) \log (\frac{B}{S}) \sqrt{\pi } T^{3/2}\\&-16 \sqrt{2} e^{-\frac{\eta ^2 (T (\sigma ^2-2 r_d+2 r_f)-2 \log (\frac{B^2}{K S}))^2}{8 T \sigma ^2}-r_d T} K S \eta  \sigma ^3 ((r_d-r_f) T+\log (\frac{B^2}{K S})) T\\&-32 \sqrt{2} e^{-\frac{\eta ^2 (T (\sigma ^2-2 r_d+2 r_f)-2 \log (\frac{B^2}{K S}))^2}{8 T \sigma ^2}-r_d T} K (r_d-r_f) S \eta  \sigma  \log (\frac{B}{S}) (T (\sigma ^2+2 r_d-2 r_f)+2 \log (\frac{B^2}{K S})) T\\&+16 B^2 e^{-\frac{\eta ^2 (T (\sigma ^2+2 r_d-2 r_f)+2 \log (\frac{B^2}{K S}))^2}{8 T \sigma ^2}-r_f T} \eta  \sigma ^3 ((r_d-r_f) T+\log (\frac{B^2}{K S})) \sqrt{2} T\\&+32 B^2 e^{-\frac{\eta ^2 (T (\sigma ^2+2 r_d-2 r_f)+2 \log (\frac{B^2}{K S}))^2}{8 T \sigma ^2}-r_f T} (r_f-r_d) \eta  \sigma  \log (\frac{B}{S}) (T (\sigma ^2-2 r_d+2 r_f)-2 \log (\frac{B^2}{K S})) \sqrt{2} T\\&-\sqrt{2} e^{-\frac{\eta ^2 (T (\sigma ^2-2 r_d+2 r_f)-2 \log (\frac{B^2}{K S}))^2}{8 T \sigma ^2}-r_d T} K S \eta ^3 \sigma  (T (\sigma ^2-2 r_d+2 r_f)-2 \log (\frac{B^2}{K S}))\times \\&\times (T (\sigma ^2+2 r_d-2 r_f)+2 \log (\frac{B^2}{K S}))^2-\sqrt{2} B^2 e^{-\frac{\eta ^2 (T (\sigma ^2+2 r_d-2 r_f)+2 \log (\frac{B^2}{K S}))^2}{8 T \sigma ^2}-r_f T} \eta ^3 \sigma\times\\ &\times  (T (\sigma ^2-2 r_d+2 r_f)-2 \log (\frac{B^2}{K S}))^2 (T (\sigma ^2+2 r_d-2 r_f)+2 \log (\frac{B^2}{K S})))
\\&\\
&\Delta(\mathcal{D})=\\&
\frac{\phi  (\frac{B}{S})^{\frac{2 (r_d-r_f)}{\sigma ^2}-1}}{2 \sigma ^2 S^2 \sqrt{T}} (2 B^2 \sqrt{T} (r_f-r_d) e^{-r_f T} \text{erf}(\frac{\eta  (2 \log (\frac{B}{S})+T (2 r_d-2 r_f+\sigma ^2))}{2 \sqrt{2} \sigma  \sqrt{T}})\\&+\sqrt{\frac{2}{\pi }} B^2 \eta  \sigma  (-\exp (-\frac{\eta ^2 (2 \log (\frac{B}{S})+T (2 r_d-2 r_f+\sigma ^2))^2}{8 \sigma ^2 T}-r_f T))-2 B^2 r_d \sqrt{T} e^{-r_f T}\\&+2 B^2 r_f \sqrt{T} e^{-r_f T}+K S \sqrt{T} e^{-r_d T} (2 r_d-2 r_f-\sigma ^2) \text{erfc}(\frac{\eta  (T (-2 r_d+2 r_f+\sigma ^2)-2 \log (\frac{B}{S}))}{2 \sqrt{2} \sigma  \sqrt{T}})\\&+\sqrt{\frac{2}{\pi }} \eta  K \sigma  S \exp (-\frac{\eta ^2 (T (-2 r_d+2 r_f+\sigma ^2)-2 \log (\frac{B}{S}))^2}{8 \sigma ^2 T}-r_d T))
\end{aligned}
\end{equation}

\begin{equation}
\nonumber
\begin{aligned}
&Vega(\mathcal{D})=\\&
\frac{\phi  (\frac{B}{S})^{\frac{2 (r_d-r_f)}{\sigma ^2}}}{4 B \sigma ^3} (8 B^2 (r_f-r_d) e^{-r_f T} \log (\frac{B}{S}) (\text{erf}(\frac{\eta  (2 \log (\frac{B}{S})+T (2 r_d-2 r_f+\sigma ^2))}{2 \sqrt{2} \sigma  \sqrt{T}})+1)\\&+\frac{\sqrt{\frac{2}{\pi }} B^2 \eta  \sigma  (T (-2 r_d+2 r_f+\sigma ^2)-2 \log (\frac{B}{S})) \exp (-\frac{\eta ^2 (2 \log (\frac{B}{S})+T (2 r_d-2 r_f+\sigma ^2))^2}{8 \sigma ^2 T}-r_f T)}{\sqrt{T}}\\&+8 K S (r_d-r_f) e^{-r_d T} \log (\frac{B}{S}) \text{erfc}(\frac{\eta  (T (-2 r_d+2 r_f+\sigma ^2)-2 \log (\frac{B}{S}))}{2 \sqrt{2} \sigma  \sqrt{T}})\\&+\frac{\sqrt{\frac{2}{\pi }} \eta  K \sigma  S (2 \log (\frac{B}{S})+T (2 r_d-2 r_f+\sigma ^2)) \exp (-\frac{\eta ^2 (T (-2 r_d+2 r_f+\sigma ^2)-2 \log (\frac{B}{S}))^2}{8 \sigma ^2 T}-r_d T)}{\sqrt{T}})
\\\\
&Vanna(\mathcal{D})=\\&
\frac{(\frac{B}{S})^{\frac{2 (r_d-r_f)}{\sigma ^2}-1} \phi}{8 \sqrt{\pi } S^2 T^{3/2} \sigma ^5}  (-8 e^{-r_d T} K \sqrt{\pi } S \sigma ^4 \text{erfc}(\frac{\eta  (T (\sigma ^2-2 r_d+2 r_f)-2 \log (\frac{B}{S}))}{2 \sqrt{2} \sqrt{T} \sigma }) T^{3/2}\\&-16 e^{-r_d T} K \sqrt{\pi } (r_d-r_f) S (-\sigma ^2+2 r_d-2 r_f) \text{erfc}(\frac{\eta  (T (\sigma ^2-2 r_d+2 r_f)-2 \log (\frac{B}{S}))}{2 \sqrt{2} \sqrt{T} \sigma }) \log (\frac{B}{S}) T^{3/2}\\&+16 B^2 e^{-r_f T} (r_d-r_f) \sigma ^2 (\text{erf}(\frac{\eta  (T (\sigma ^2+2 r_d-2 r_f)+2 \log (\frac{B}{S}))}{2 \sqrt{2} \sqrt{T} \sigma })+1) \sqrt{\pi } T^{3/2}\\&+8 e^{-r_d T} K S \sigma ^2 (\sigma ^2-2 r_d+2 r_f) \text{erfc}(\frac{\eta  (T (\sigma ^2-2 r_d+2 r_f)-2 \log (\frac{B}{S}))}{2 \sqrt{2} \sqrt{T} \sigma }) \sqrt{\pi } T^{3/2}\\&+16 B^2 e^{-r_f T} (r_f-r_d) \sigma ^2 (\text{erf}(\frac{\eta  (T (\sigma ^2+2 r_d-2 r_f)+2 \log (\frac{B}{S}))}{2 \sqrt{2} \sqrt{T} \sigma })+1) \log (\frac{B}{S}) \sqrt{\pi } T^{3/2}\\&+16 B^2 e^{-r_f T} (r_d-r_f) (\sigma ^2+2 r_d-2 r_f) (\text{erf}(\frac{\eta  (T (\sigma ^2+2 r_d-2 r_f)+2 \log (\frac{B}{S}))}{2 \sqrt{2} \sqrt{T} \sigma })+1) \log (\frac{B}{S})\times\\&\times \sqrt{\pi } T^{3/2}-4 \sqrt{2} e^{-\frac{\eta ^2 (T (\sigma ^2-2 r_d+2 r_f)-2 \log (\frac{B}{S}))^2}{8 T \sigma ^2}-r_d T} K S \eta  \sigma ^3 T-16 \sqrt{2} e^{-\frac{\eta ^2 (T (\sigma ^2-2 r_d+2 r_f)-2 \log (\frac{B}{S}))^2}{8 T \sigma ^2}-r_d T}K \times\\&\times (r_d-r_f) S \eta  \sigma  \log (\frac{B}{S}) T-2 \sqrt{2} B^2 e^{-\frac{\eta ^2 (T (\sigma ^2+2 r_d-2 r_f)+2 \log (\frac{B}{S}))^2}{8 T \sigma ^2}-r_f T} \eta  \sigma  (\sigma ^2+2 r_d-2 r_f)\times\\&\times (T (\sigma ^2-2 r_d+2 r_f)-2 \log (\frac{B}{S})) T+4 B^2 e^{-\frac{\eta ^2 (T (\sigma ^2+2 r_d-2 r_f)+2 \log (\frac{B}{S}))^2}{8 T \sigma ^2}-r_f T} \eta  \sigma ^3 \sqrt{2} T\\&+16 B^2 e^{-\frac{\eta ^2 (T (\sigma ^2+2 r_d-2 r_f)+2 \log (\frac{B}{S}))^2}{8 T \sigma ^2}-r_f T} (r_d-r_f) \eta  \sigma  \log (\frac{B}{S}) \sqrt{2} T+2 e^{-\frac{\eta ^2 (T (\sigma ^2-2 r_d+2 r_f)-2 \log (\frac{B}{S}))^2}{8 T \sigma ^2}-r_d T}\times \\&\times K S \eta  \sigma  (\sigma ^2-2 r_d+2 r_f) (T (\sigma ^2+2 r_d-2 r_f)+2 \log (\frac{B}{S})) \sqrt{2} T\\&+2 B^2 e^{-\frac{\eta ^2 (T (\sigma ^2+2 r_d-2 r_f)+2 \log (\frac{B}{S}))^2}{8 T \sigma ^2}-r_f T} \eta  \sigma ^3 (T (\sigma ^2-2 r_d+2 r_f)-2 \log (\frac{B}{S})) \sqrt{2} T\\&-\sqrt{2} e^{-\frac{\eta ^2 (T (\sigma ^2-2 r_d+2 r_f)-2 \log (\frac{B}{S}))^2}{8 T \sigma ^2}-r_d T} K S \eta ^3 \sigma  ((\sigma ^4-4 r_d^2-4 r_f^2+8 r_d r_f) T^2+8 (r_f-r_d) \log (\frac{B}{S}) T\\&-4 \log ^2(\frac{B}{S}))+B^2 e^{-\frac{\eta ^2 (T (\sigma ^2+2 r_d-2 r_f)+2 \log (\frac{B}{S}))^2}{8 T \sigma ^2}-r_f T} \eta ^3 \sigma  ((\sigma ^4-4 r_d^2-4 r_f^2\\&+8 r_d r_f) T^2+8 (r_f-r_d) \log (\frac{B}{S}) T-4 \log ^2(\frac{B}{S})) \sqrt{2})
\end{aligned}
\end{equation}
\begin{equation}
\nonumber
\begin{aligned}
&Volga(\mathcal{D})=\\&
\frac{(\frac{B}{S})^{\frac{2 (r_d-r_f)}{\sigma ^2}} \phi}{16 B \sqrt{\pi } T^{3/2} \sigma ^6}  (-128 e^{-r_d T} K \sqrt{\pi } (r_d-r_f)^2 S \text{erfc}(\frac{\eta  (T (\sigma ^2-2 r_d+2 r_f)-2 \log (\frac{B}{S}))}{2 \sqrt{2} \sqrt{T} \sigma }) \log ^2(\frac{B}{S}) T^{3/2}\\&-96 e^{-r_d T} K \sqrt{\pi } (r_d-r_f) S \sigma ^2 \text{erfc}(\frac{\eta  (T (\sigma ^2-2 r_d+2 r_f)-2 \log (\frac{B}{S}))}{2 \sqrt{2} \sqrt{T} \sigma }) \log (\frac{B}{S}) T^{3/2}\\&+128 B^2 e^{-r_f T} (r_d-r_f)^2 (\text{erf}(\frac{\eta  (T (\sigma ^2+2 r_d-2 r_f)+2 \log (\frac{B}{S}))}{2 \sqrt{2} \sqrt{T} \sigma })+1) \log ^2(\frac{B}{S}) \sqrt{\pi } T^{3/2}\\&+96 B^2 e^{-r_f T} (r_d-r_f) \sigma ^2 (\text{erf}(\frac{\eta  (T (\sigma ^2+2 r_d-2 r_f)+2 \log (\frac{B}{S}))}{2 \sqrt{2} \sqrt{T} \sigma })+1) \log (\frac{B}{S}) \sqrt{\pi } T^{3/2}\\&-16 \sqrt{2} e^{-\frac{\eta ^2 (T (\sigma ^2-2 r_d+2 r_f)-2 \log (\frac{B}{S}))^2}{8 T \sigma ^2}-r_d T} K S \eta  \sigma ^3 ((r_d-r_f) T+\log (\frac{B}{S})) T\\&-32 \sqrt{2} e^{-\frac{\eta ^2 (T (\sigma ^2-2 r_d+2 r_f)-2 \log (\frac{B}{S}))^2}{8 T \sigma ^2}-r_d T} K (r_d-r_f) S \eta  \sigma  \log (\frac{B}{S}) (T (\sigma ^2+2 r_d-2 r_f)\\&+2 \log (\frac{B}{S})) T+16 B^2 e^{-\frac{\eta ^2 (T (\sigma ^2+2 r_d-2 r_f)+2 \log (\frac{B}{S}))^2}{8 T \sigma ^2}-r_f T} \eta  \sigma ^3 ((r_d-r_f) T+\log (\frac{B}{S})) \sqrt{2} T\\&+32 B^2 e^{-\frac{\eta ^2 (T (\sigma ^2+2 r_d-2 r_f)+2 \log (\frac{B}{S}))^2}{8 T \sigma ^2}-r_f T} (r_f-r_d) \eta  \sigma  \log (\frac{B}{S}) (T (\sigma ^2-2 r_d+2 r_f)-2 \log (\frac{B}{S})) \sqrt{2} T\\&-\sqrt{2} e^{-\frac{\eta ^2 (T (\sigma ^2-2 r_d+2 r_f)-2 \log (\frac{B}{S}))^2}{8 T \sigma ^2}-r_d T} K S \eta ^3 \sigma  (T (\sigma ^2-2 r_d+2 r_f)-2 \log (\frac{B}{S})) (T (\sigma ^2+2 r_d-2 r_f)\\&+2 \log (\frac{B}{S}))^2-\sqrt{2} B^2 e^{-\frac{\eta ^2 (T (\sigma ^2+2 r_d-2 r_f)+2 \log (\frac{B}{S}))^2}{8 T \sigma ^2}-r_f T} \eta ^3 \sigma  (T (\sigma ^2-2 r_d+2 r_f)\\&-2 \log (\frac{B}{S}))^2 (T (\sigma ^2+2 r_d-2 r_f)+2 \log (\frac{B}{S})))
\end{aligned}
\end{equation}
\section{The Vanna-Volga Method}
The Vanna-Volga method, also known as the trader's rule of thumb, is a pricing technique to determine the cost of risk managing the volatility risk of highly volatility options, such as, FX-options. This cost is then added to the theoretical value predicted by the BS model. Several papers deals with the Vanna-Volga method and its extensions (see for instance \cite{bossens,Wystup,Mercurio}), nevertheless, we present in this section a unified treatment of the method, as well as some extensions of it. The Vanna-Volga (VV) method is commonly used in FX options markets, where three main volatility quotes are typically available for a given maturity
\begin{enumerate}
	\item The $0\Delta$ straddle,
	\item  the $25\Delta$ risk reversal,
	\item and the $25\Delta$ Vega weighted butterfly
\end{enumerate}
however, the choice of those quotes are based on a liquidity matter. Our point of view will be as the one explained in \cite{Shkolkikov}, in the sense that a general VV-method will be derived from the Itô formula, which is applicable to any kind of option. From know on we will assume  that for each $t$, the underlying $S$ is  observable and follows an standard BS -log normal process with volatility $\sigma_t$, therefore, $\sigma_t$ is a random variable, obtained from market information, and will be known as a fair value implied volatility (see \cite{Shkolkikov}). Additionally, $\sigma_{t}$ will be assumed to be a log-normal diffusion process, and we will fix three listed pivot European vanilla calls $C_i$ at strikes $K_i$, maturing at $T$ or later, and traded at public markets with quotes available all the time.  \\\

Let $O$ be an European option (not necessarily an FX derivative) with maturity $T$, which depends on two random variables: the spot price of the underlying asset and the volatility. The goal is to find the market price $O^{Mk}$ of $O$ as the Black-Scholes price of the option $O^{BS}$ which is computed using the fair value implied volatility, plus a correction taking into account a potential dependence of the implied volatility of pivot market prices $C_i^{Mk}$ from strikes $K_i$.  Following \cite{Shkolkikov,Mercurio}, we will  replicate $O^{BS}$ by constructing a risk-neutral portfolio $\Pi^{BS}$ going long $1$ unit of $O$,  being going short $\Delta_{t}$ units of $S$, and going short $x_i$ units of each pivot.  Using the Itô lemma,

\begin{equation}
\begin{aligned}d O^{BS}(t)&=\frac{\partial O^{BS}}{\partial t} dt+\frac{\partial O^{BS}}{\partial S} dS_t+\frac{\partial O^{BS}}{\partial \sigma} d\sigma_t+\\ \\&+\frac{1}{2}(\frac{\partial^2 O^{BS}}{\partial S^2} (dS_t)^2+\frac{\partial^2 O^{BS}}{\partial \sigma^2} (d\sigma_t)^2+2\frac{\partial^2 O^{BS}}{\partial S\partial \sigma} dS_td\sigma_t)
\end{aligned}\end{equation}

Therefore, the value variation of the portfolio $\Pi^{BS}$ is
\begin{equation}\label{port-hed}
\begin{aligned}
d\Pi^{BS}&=d(O^{BS}(t)-\Delta_t S_t-\sum_{i=1}^{3}x_{i}C_{i}^{BS}(t))=\\&\left[ \frac{\partial O^{BS}}{\partial t} - \sum_i x_i \frac{\partial C_{i}^{BS}}{\partial t}  \right] dt   + \left[ \frac{\partial O^{BS}}{\partial S} - \Delta_t - \sum_i x_i \frac{\partial C_{i}^{BS}}{\partial S}  \right] dS_t  +\\ \\&+ \left[ \frac{\partial O^{BS}}{\partial \sigma} - \sum_i x_i \frac{\partial C_{i}^{BS}}{\partial \sigma}  \right] d\sigma_t  + \frac{1}{2} \left[ \frac{\partial^2 O^{BS}}{\partial S^2} - \sum_i x_i \frac{\partial^2 C_{i}^{BS}}{\partial S^2}  \right] (dS_t)^2  +\\ \\&+ \frac{1}{2} \left[ \frac{\partial^2 O^{BS}}{\partial \sigma^2} - \sum_i x_i \frac{\partial^2 C_{i}^{BS}}{\partial \sigma^2}  \right] (d\sigma_t)^2  + \left[ \frac{\partial^2 O^{BS}}{\partial S \partial \sigma} - \sum_i x_i \frac{\partial^2 C_{i}^{BS}}{\partial S \partial \sigma}  \right] dS_t d\sigma_t
\end{aligned}
\end{equation}

It is possible to select $x_1,x_2,x_3$ to zero out the terms $dS_t, d\sigma_t, (d\sigma_t)^2$ and $dS_td\sigma_t$, and matching \begin{equation}\Delta_t=\frac{\partial(O^{BS}-\sum_{i=1}^{3}x_{i}C_{i}^{BS})}{\partial{S}}(t,S_t)
\end{equation} so, equation \eqref{port-hed} is transformed into
\begin{equation}\label{port-he2}
d\Pi^{BS}=\left[ \frac{\partial O^{BS}}{\partial t} - \sum_i x_i \frac{\partial C_{i}^{BS}}{\partial t}  \right] dt +\left[ \frac{\partial^2 O^{BS}}{\partial S^2} - \sum_i x_i \frac{\partial^2 C_{i}^{BS}}{\partial S^2}  \right] (dS_t)^2
\end{equation}
Now,
\begin{equation}\label{sto}
\begin{aligned}
(dS_t)^2&=(\mu S_t dt+\sigma_t S_tdW_t)^2\\
\\&=\mu^2 S_t^2 (dt)^2+2\mu\sigma_tS_t^2 dtdW_t+\sigma^2_tS_t^2(dW_t)^2=\sigma^2_t S_t^2 dt
\end{aligned}
\end{equation}
where the last equations follows from the fact that is possible to substitute $dt$ for $dW_t^2$ due to the quadratic variation of a Weiner process, and $dt^2$ and $dt dB_t$ can be set as zero due to the rules of Stochastic calculus. For a complete account of stochastic calculus we refer the reader to \cite{shreve}. Applying equation \eqref{sto}, the final expression of the value variation of portfolio $\Pi^{BS}$ is
\begin{equation}
\begin{aligned}
d\Pi^{BS}=\left[(\frac{\partial O^{BS}}{\partial t} - \sum_i x_i \frac{\partial C_{i}^{BS}}{\partial t}) +\frac{\sigma_t^2S_t^2}{2}( \frac{\partial^2 O^{BS}}{\partial S^2} - \sum_i x_i \frac{\partial^2 C_{i}^{BS}}{\partial S^2})  \right] (d_t)
\end{aligned}
\end{equation}

Finally, the assumption that both $O^{BS}$ and $C_{i}^{BS}$ satisfy the Black-Scholes PDE-equation implies that $\Pi^{BS}$ is a self-financing portfolio, that is
\begin{equation}d\Pi^{BS}=r\Pi^{BS}dt
\end{equation}

If $O$ is an FX-option, the above equation is written as
\begin{equation}d\Pi^{BS}=r_d\Pi^{BS}dt
\end{equation}

This framework is more suitable that the one developed in \cite{Mercurio}, because it is applicable to all kind of options, in contrast with the aforementioned paper, where only European vanillas FX options are taking into consideration.
The parameters $x_i$ are obtained by solving the following system of equations

\begin{equation}
\begin{aligned}
\frac{\partial O^{BS}}{\partial \sigma} &= \sum_i x_i \frac{\partial C^{BS}_i}{\partial \sigma} \\
\frac{\partial^2 O^{BS}}{\partial \sigma^2} &= \sum_i x_i \frac{\partial^2 C^{BS}_i}{\partial \sigma^2} \\
\frac{\partial^2 O^{BS}}{\partial S \partial \sigma} &= \sum_i x_i \frac{\partial^2 C^{BS}_i}{\partial S \partial \sigma}
\end{aligned}
\end{equation}

or, equivalently
\renewcommand{\arraystretch}{1}
\begin{equation}\label{system}
\mathbb{A}= \mathbb{V}\cdot \begin{pmatrix}
x_1 \\
x_2\\
x_3
\end{pmatrix}
\end{equation}
where

\begin{equation}
\mathbb{A}:=\begin{pmatrix}
Vega(O^{BS})  \\[0.2cm]
Vanna(O^{BS}) \\[0.2cm]
Volga(O^{BS})
\end{pmatrix}\hspace{0.7cm}\mathbb{V}:=\begin{pmatrix}
Vega(C_1^{BS}) & Vega(C_2^{BS}) & Vega(C_3^{BS})  \\[0.2cm]
Vanna(C_1^{BS}) & Vanna(C_2^{BS}) & Vanna(C_3^{BS}) \\[0.2cm]
Volga(C_1^{BS}) & Volga(C_2^{BS}) & Volga(C_3^{BS}) \\
\end{pmatrix}
\end{equation}
We have proved the following.
\begin{theorem}
	If the underlying $S$ follows a Black-Scholes process with a stochastic, but strike-independent implied volatility, any option contract $O$ expiring at $T$, can be locally perfectly hedge by a portfolio of one unit of $O$, $\Delta$ units of the underlying, and $x_i$ units of pivot $C_i$, $i=1,...,3$, expiring at $T$ or later, where the parameters $x_1,x_2$ and $x_3$ can be found from \eqref{system}. Due to the put-call parity, any replacement of $C_i$ with a put $P_i$ with strike $K_i$, changes $\Delta$ but does not change $x_i$.\end{theorem}
In the case we don't have enough market information, it is possible to obtain a self financing portfolio adding extra hypotheses as it is proved in \cite{Shkolkikov}.
\begin{theorem}\cite[Proposition 2]{Shkolkikov}If $S_t$
	follows a geometric Brownian motion with stochastic but strike-independent implied volatility, there exists a unique self-financing portfolio
	\begin{equation}
	\Pi^{Mk}=O^{Mk}-\Delta^{Mk}dS-\sum_{i=1}^{3}x_i C_i^{Mk}
	\end{equation}
	satisfying the property $\Pi^{Mkt}=\Pi^{BS}$ at any time $0\leq t\leq T$. This implies that for any contract $O$, the Vanna-Volga price is given by
	\begin{equation}O^{MK}_{VV}=O^{BS}+\sum_{i=1}^{3}x_{i}(C_{i}^{MK}-C_{i}^{BS})
	\end{equation}
\end{theorem}
\subsection{Vanna-Volga in FX markets}
From now on we will focus our attention on FX markets. Due to its high liquidity, the  benchmark securities $C_i$ that we will be using are a delta neutral straddle, a $25\Delta$ risk reversal and a $25\Delta$ vega weighted butterfly that will be denoted, following the nomenclature used in \cite{bossens,Mercurio} as
\begin{equation}\label{products}
\begin{aligned}
&\text{ATM}=\frac{1}{2}\text{Straddle}(K_{ATM})=\frac{1}{2}(C(K_{ATM},\sigma_{ATM})+P((K_{ATM},\sigma_{ATM}))\\
&\text{BF}=\frac{1}{2}((C(K_{c},\sigma_{K_c})+P(K_{p},\sigma_{p})-(C(K_{ATM},\sigma_{ATM})+P(K_{ATM},\sigma_{ATM})))\\
&\text{RR}=C(K_c,\sigma(K_c))-P(K_p,\sigma(K_p))
\end{aligned}
\end{equation}
where $K_{ATM}$ denotes the $ATM$ strike, $K_{c/p}$ the $25\Delta$ call/put strikes are obtained by solving the equations
\begin{equation}
\Delta_{Call}(K_{c},\sigma_{ATM})=0.25\hspace{1cm} \Delta_{Put}(K_{p},\sigma_{ATM})=-0.25
\end{equation}
and $\sigma(K_{c/p})$ the corresponding volatilities evaluated from the smile surface. For the sake of consistency, let us explain how to obtain $\sigma(K_{C/P})$ in FX-markets. Generally speaking, brokers quote volatilities expressed as functions of $\Delta$, instead of given direct prices of the instruments described in equation \eqref{products}. For instance, a $25\Delta_{Call}$-volatility refers to the volatility at the strike $K_{c}$ that satisfies
\begin{equation}
\Delta_{Call}(K_c,\sigma(K_c))=0.25
\end{equation}
By considering the instruments given in \eqref{products}, we observe that the Black-Scholes price and the Market price of an ATM straddle are identical, so we obtain that
\begin{equation}
O^{VV}=O^{BS}+x_2(RR^{Mk}-RR^{BS})+x_3(BF^{Mk}-BF^{BS})
\end{equation}
with $x_2$ and $x_3$ obtained through the linear system \eqref{system} which can be easily computed since in the previous section we have obtained closed formulas for the Vega, Vanna and Volga for any kind of single/double barrier option. Let us remark here that the Black-Scholes price of $O$, as well as the Greeks, are computed with the ATM volatility.

\end{document}